\documentclass[sn-mathphys,Numbered]{sn-jnl}% Math and 
\usepackage{graphicx}%
\usepackage{multirow}%
\usepackage{amsmath,amssymb,amsfonts}%
\usepackage{amsthm}%
\usepackage{mathrsfs}%
\usepackage[title]{appendix}%
\usepackage{xcolor}%
\usepackage{textcomp}%
\usepackage{manyfoot}%
\usepackage{booktabs}%
\usepackage{algorithm}%
\usepackage{algorithmicx}%
\usepackage{algpseudocode}%
\usepackage{listings}%
\begin{document}

\title[Article Title]{RFSoC Gen3-Based Software-Defined Radio Characterization for the Readout System of Low-Temperature Bolometers}


%% Authors PhD Students
\author*[1,2,4,5,6]{\fnm{M. E.} \sur{García Redondo}}\email{manuel.garcia@iteda.cnea.gov.ar}
\author[2]{\fnm{T.} \sur{Muscheid}}
\author[2]{\fnm{R.} \sur{Gartmann}}
\author[1,2,5]{\fnm{J. M.} \sur{Salum}}
\author[1,2,4,5,6]{\fnm{L. P.} \sur{Ferreyro}}
\author[1,3,5]{\fnm{N. A.} \sur{Müller}}
\author[1,3,4,5]{\fnm{J. D.} \sur{Bonilla-Neira}}
\author[1,3,4,5,6]{\fnm{J. M.} \sur{Geria}}
\author[3,5,7]{\fnm{J. J.} \sur{Bonaparte}}

%% Authors Researchers
\author[1,6,7]{\fnm{A.} \sur{Almela}}
\author[2]{\fnm{L. E.} \sur{Ardila-Perez}} % https://orcid.org/0000-0002-7485-8267
\author[1,4,6,7]{\fnm{M. R.} \sur{Hampel}}
\author[1,4,6,7]{\fnm{A. E.} \sur{Fuster}}

%% Group Heads
\author[1,4,5,7]{\fnm{M.} \sur{Platino}}
\author[2]{\fnm{O.} \sur{Sander}}
\author[2]{\fnm{M.} \sur{Weber}}
\author[1]{\fnm{A.} \sur{Etchegoyen}}

\affil[1]{\orgname{Instituto de Tecnologías en Detección y Astropartículas (ITeDA)}, \country{Argentina}}
\affil[2]{\orgname{Institute for Data Processing and Electronics (IPE)},  \orgname{Karlsruhe Institute of Technology (KIT)}, \country{Germany}}
\affil[3]{\orgname{Institute of Micro- and Nanoelectronic Systems (IMS)},  \orgname{Karlsruhe Institute of Technology (KIT)}, \country{Germany}}
\affil[4]{\orgname{Consejo Nacional de Investigaciones
Científicas y Técnicas (CONICET)}, \country{Argentina}}
\affil[5]{\orgname{Universidad Nacional de San Martín (UNSAM)}, \country{Argentina}}
\affil[6]{\orgname{Universidad Tecnológica Nacional (UTN)}, \country{Argentina}}
\affil[7]{\orgname{Comisión Nacional de Energía Atómica (CNEA)}, \country{Argentina}}


\abstract{This work reports the performance evaluation of an SDR readout system based on the latest generation (Gen3) of AMD's Radio Frequency System-on-Chip (RFSoC) processing platform, which integrates a full-stack processing system and a powerful FPGA with up to 32 high-speed and high-resolution 14-bit Digital-to-Analog Converters (DACs) and 14-bit Analog-to-Digital Converters (ADCs).
The proposed readout system uses a previously developed multi-band, double-conversion IQ RF-mixing board targeting a multiplexing factor of approximately 1,000 bolometers in a bandwidth between 4 and 8 GHz, in line with state-of-the-art microwave SQUID multiplexers (\textmu MUX). 
The characterization of the system was performed in two stages, under the conditions typically imposed by the multiplexer and the cold readout circuit. 
First, in transmission, showing that noise and spurious levels of the generated tones are close to the values imposed by the cold readout. 
Second, in RF loopback, presenting noise values better than -100 dBc/Hz totally in agreement with the state-of-the-art readout systems. 
It was demonstrated that the RFSoC Gen3 device is a suitable enabling technology for the next generation of superconducting detector readout systems, reducing system complexity, increasing system integration, and achieving these goals without performance degradation.}


\maketitle

%%% Introduction Section
\section{Introduction}\label{intro}
The next generation of Cosmic Microwave Background (CMB) B-mode polarization telescopes require densely populated focal planes with thousands of ultra-sensitive bolometers operating at cryogenic temperatures to achieve the required sensitivity\,\cite{Abitbol2017}. However, reading out a large number of detectors such as Transition Edge Sensors (TES)\,\cite{Hubmayr_2018} or the recently proposed Magnetic Microbolometers (MMBs)\,\cite{Geria2023} at temperatures below one Kelvin imposes significant technical challenges to the cryogenic readout systems. In the last decade, a technique called Microwave Superconducting Quantum Interference Device (SQUID) Multiplexing ($\mu$MUXing) has become predominant because it is able to achieve multiplexing factors in excess of 1000 while maintaining the readout noise subdominant to the intrinsic detector noise\,\cite{Dober2020}. This scheme encodes the detector signals in the resonance frequencies of multiple GHz-frequency superconducting resonators coupled to a common feed-line. Therefore, recovering the signals from each detector is straightforward, requiring only monitoring the resonant frequencies. Despite the aforementioned benefits, it imposes stringent requirements for generating and acquiring high-purity broadband microwave signals on the warm electronics and real-time processing without degrading the performance imposed by the cold multiplexing circuit\,\cite{Rao_2020}.

This task is traditionally tackled using FPGA-based Software-Defined Radio (SDR) systems offering high customizability, large I/O bandwidths and real-time processing. The FPGA drives a set of high-speed data converters along with analog mixers and other Radio-Frequency (RF) components to translate these signals to the required operating band\,\cite{Yu2022}. Nowadays, technological advances in telecommunications have led to developments such as the AMD's Radio Frequency System-on-Chip (RFSoC), which combines a high-performance heterogeneous processing platform with several embedded high-speed digitizers ($>$1\,GSPS) into a System-on-Chip (SoC)\,\cite{rfsoc_over}. This kind of SoC significantly increases the total bandwidth capabilities while simultaneously reducing system size, weight, and power consumption in comparison to traditional systems.

We seek to improve the state-of-the-art of the frequency-multiplexed SDR detector readout systems by migrating to an RFSoC platform and making use of the extended characteristics of the newest Gen3 devices, such as an improved number of data converters with faster sampling rates and an augmented number of bits\,\cite{rfsoc_ac}. We present the noise performance characterization of a custom RFSoC Gen3-based SDR readout system for low-temperature bolometers multiplexed in the frequency domain by means of the Microwave SQUID Multiplexer ($\mu$MUX).

%%% System Overview Section
\section{System Overview}\label{sysover}
\begin{figure}[h]%
\centering
\includegraphics[width=1\textwidth]{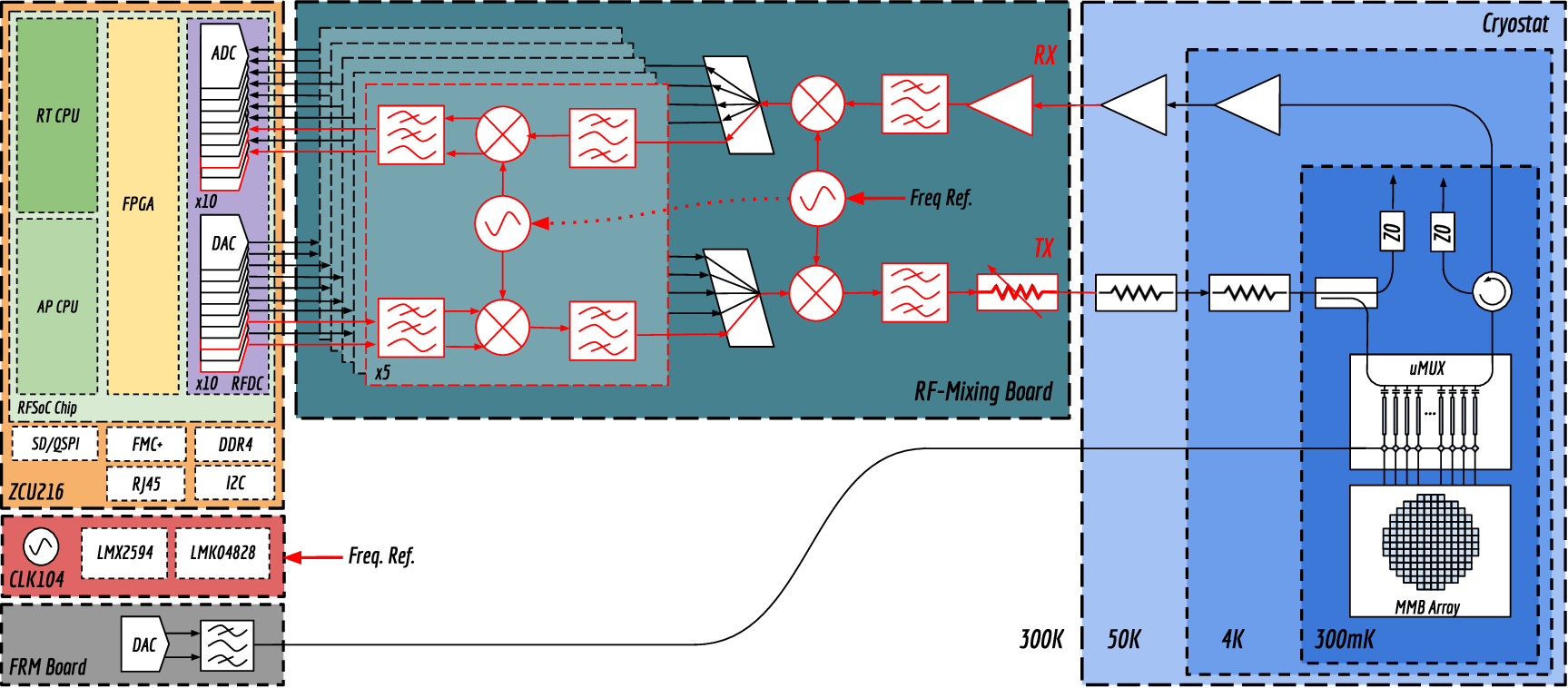}
\caption{RFSoC SDR readout system built around the ZCU216 Evaluation Kit, including an external CLK104 RF clock add-on board and a flux-ramp generation boards (left). Multi-band IQ RF-mixing board (center). Cold multiplexing system and bolometer array (right).}
\label{sdrscheme}
\end{figure}

The proposed system shown in Figure\,\ref{sdrscheme} is built around the Zynq™ UltraScale+™ RFSoC ZCU216 Evaluation Kit\,\cite{zcu216}. It adapts hardware and firmware architecture developed for the The Electron Capture in \textsuperscript{163}Ho experiment (ECHo)\,\cite{sander19} to the RFSoC Gen3 family of devices. The system is targeting the multiplexed readout of 1000 low-temperature bolometers in the bandwidth between 4 and 8\,GHz, in line with state-of-the-art \textmu MUXing systems\,\cite{Dober2020}.

The hardware architecture is based on a multi-band, double conversion, zero-intermediate frequency RF-mixing board, which divides the total bandwidth into five 800\,MHz complex base-bands\,\cite{sander19}. As a consequence, ten DACs and ten ADCs are required to cover the total bandwidth. The adopted sampling rate is 1\,GSPS due to the excellent performance in output power, Spurious-Free Dynamic Range (SFDR) and Signal-to-Noise Ratio (SNR) in agreement with the datasheet and previously used converters\,\cite{rfsoc_ac,Gartmann2022}. Ultra-stable low-noise clock generation is achieved using the CLK104 RF clock add-on card\,\cite{clk104} connected to an external 10\,MHz frequency reference fanned out through the different boards. Because the multi-band RF-mixing board is still under development, an available single complex base-band prototype version connected to the RFSoC converters through the XM655 add-on card was used for the characterization\,\cite{Gartmann2022}. This prototype RF-mixing board comprises the blocks highlighted in red in Figure\,\ref{sdrscheme} and allows us to test a single sub-band using one DAC pair and one ADC pair. Due to the RF-mixing board modular structure, the characterization of a single sub-band is representative of the final multi-band system performance.

Regarding the firmware, the RFSoC Gen3 was integrated into our custom build system based on Yocto and most of the blocks previously developed were reused \,\cite{Karcher2020,Ferreyro2023}. The main contribution of this work was the integration and configuration of the new data converters into the firmware design by means of the Radio Frequency Data Converter RFDC™ LogiCORE™ IP\,\cite{rfdc}. Since the same Zynq™ UltraScale+™ device family is used for the reported development, it also allowed to keep the same Processing System (PS) software, a huge advantage for fast development.

%%% System Performance Section
\section{System Performance}\label{performance}
The cold multiplexing system depicted in Figure\,\ref{sdrscheme} encodes the detector signals into phase and amplitude of the probing tones. Therefore, the noise present at these coordinates is indistinguishable from the detected signals\,\cite{ahrensthesis}. The scope of this work is to quantify the noise degradation due to the SDR system with respect to the cryogenic Low-Noise Amplifier (LNA) noise within the range of possible frequencies adopted by the flux-ramp modulation for bolometric applications and assuming a phase domain readout\,\cite{Mates2012,Salum2023}.

\begin{figure}[H]%
\centering
\includegraphics[width=1\textwidth]{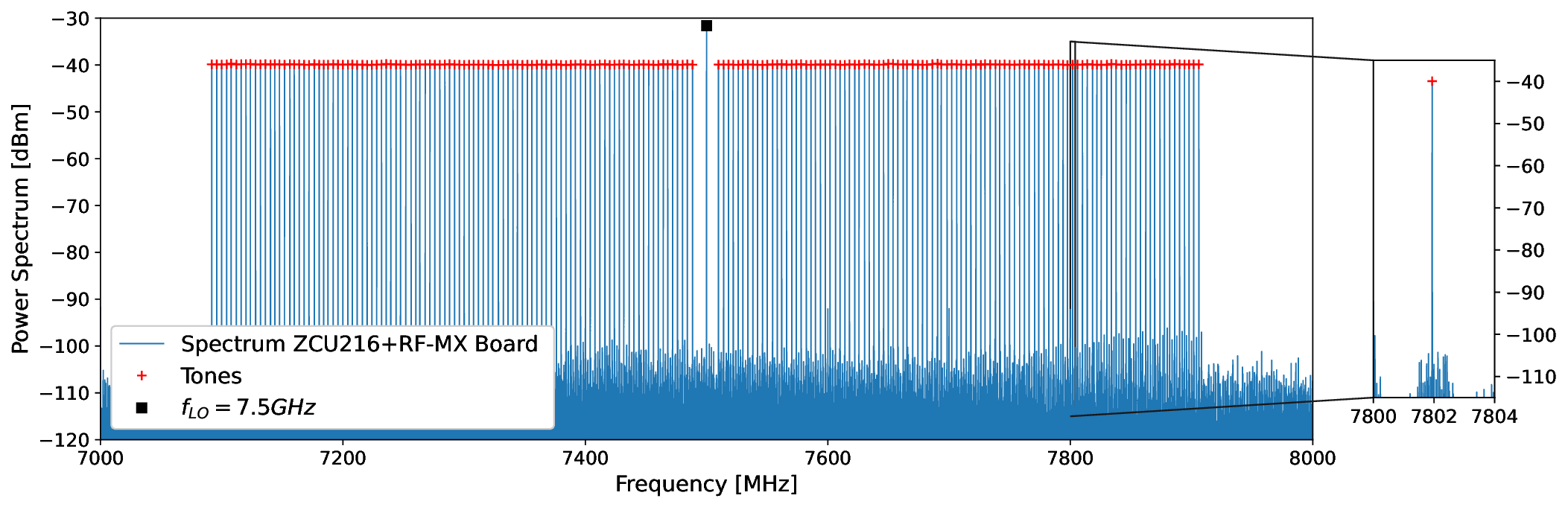}
\caption{Power spectrum of 200 fixed tones generated by the RFSoC-based SDR system proposed in this work using random frequency spacing ($\mu=4$\,MHz spacing and $\sigma=200$\,kHz deviation) centered at 7.5\,GHz with -40\,dBm power per tone. The inset shows a close-in-look at one tone taken as example to see the close-in spurious signals produced by the SDR system.}\label{tones_tx}
\end{figure}

First, the signal generation performance was characterized seeking to ensure the quality of the tones required for monitoring the $\mu$MUX channels. For this, a number of N = 200 tones centered at 7.5\,GHz were generated with a Gaussian frequency distribution with $\mu=4$\,MHz spacing and $\sigma=200$\,kHz deviation. This emulates the resonance frequency distribution of a real multiplexer given by variations in the manufacturing process and allows for the identification of inter-modulation products generated within the readout bandwidth. Comb generation was performed by continuously reproducing the complex waveform stored in Block RAM (BRAM) at a rate of 1\,GSPS by the DACs. It has a memory depth of $2^{17}$ complex samples resulting in a frequency resolution of  $\Delta f\approx7.6$\,kHz which is sufficient for frequency placement in bolometric applications where the resonator bandwidth is BW$\approx$200\,kHz~\cite{Dober2020}. Then, the base-band signal was up-converted to 7.5\,GHz by the RF-mixing board. Figure\,\ref{tones_tx} shows the generated frequency comb at the RF-mixing board transmitter output (Tx). The selected tone power at Tx port is -40\,dBm considering an optimum readout power of -75\,dBm at the $\mu$MUX input port and 35\,dB of cold attenuation\,\cite{Schuster_2023}. In order to utilize the DACs dynamic range efficiently, several Peak-to-Average Power Ratio (PAPR) minimization methods were evaluated\,\cite{Shibasaki} but none of them performed significantly better than the random phases method with an average PAPR $\approx$ 12\,dB in case of non-uniformly spaced tones.

\begin{figure}[h]%
\centering
\includegraphics[width=0.9\textwidth]{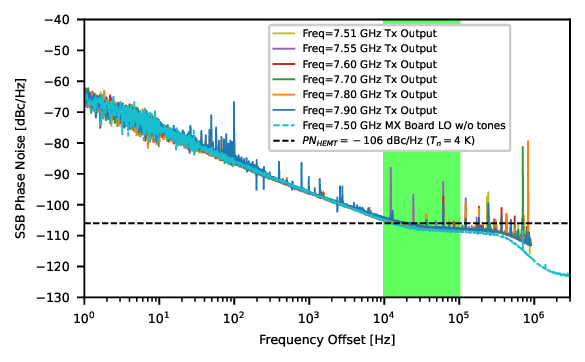}
\caption{SSB phase noise for selected tones at transmitter output (Tx). Green and black dashed lines are the up-conversion local oscillator and cryogenic low-noise amplifier phase noise profiles, respectively. The shaded green region represents the possible region for Flux-Ramp Modulation (FRM) frequencies.}\label{phase_tx}
\end{figure}

The transmitter output (Tx) was connected to a R\&S FSWP50 phase noise analyzer, where six reference tones were measured regarding their phase noise. This selected subset is representative of the quality of all generated tones. The results of these measurements are presented in Figure\,\ref{phase_tx}. The shaded green region represents the frequency band where the detector signal flux-ramp modulation will be. Typically between 10\,kHz and 100\,kHz for bolometric applications\,\cite{Dober2020,Yu2022}. The dashed cyan line represents the phase noise present in the local oscillator of the RF-mixing board without tone generation. It can be seen that this local oscillator imposes the shape of the stimulation phase noise. The phase noise profile in the band of interest is dominated by Voltage-Controlled Oscillator (VCO) white phase noise and spurious signals, while the quantization noise is negligible. This value is compared with the estimated cryogenic LNA noise level represented in the dashed black line in Figure\,\ref{phase_tx}. LNA Single-Sideband (SSB) phase noise $\mathscr{L}(f)$ can be calculated according
to \cite{rubiola2010phase}

\begin{equation}
    \mathscr{L}(f)=\frac{k_BT_n}{2 |S_{21}^{min}|^2 P_{r}}
\end{equation}

where $k_B$ is the Boltzmann constant, $T_n$ is the LNA equivalent noise temperature, $S_{21}^{min}$ is the resonance depth and $P_r$ the tone power at \textmu MUX input. Assuming an LNA equivalent noise temperature of $T_{n}\approx4$\,K, resonance depth of $S_{21}^{min} \equiv -15$\,dB and $P_r\approx-75$\,dBm as mentioned earlier, the minimum Single-Sideband (SSB) phase noise density at the multiplexer output is $\mathscr{L}(f)\approx$ -106\,dBc/Hz. Therefore, a maximum degradation of 3\,dB with respect to the contribution of the cryogenic LNA is expected due to the phase noise present in the probing tones.

Later, once the spectral purity of the generated tones had been characterized, we evaluated the degradation of the tones due to the receiver path. A loop between the transmitter (Tx) and receiver (Rx) ports of the RF-mixing board was created through a $\approx$ 10\,m long cable with around 15\,dB attenuation, emulating the typical input-output attenuation of the cryogenic RF circuit at resonance~\cite{Rao_2020,Rantwijk}. This is consistent with 15\,dB of net attenuation resulting from the combination of 35\,dB of cold attenuation, a typical LNA gain of 35\,dB and -15\,dB resonance depth. Considering that the tones were generated with random phases and the fact that in a real measurement both the RF components and the detector signals will lead to a pseudo-randomization of the phases, a PARP $\approx$ 12\,dB at the input of the ADCs was considered. Hence, the attenuation of the Rx path was adjusted to satisfy $P_{tone}\approx-40$\,dBFS for optimum ADC SFDR.

\begin{figure}[h]%
\centering
\includegraphics[width=0.9\textwidth]{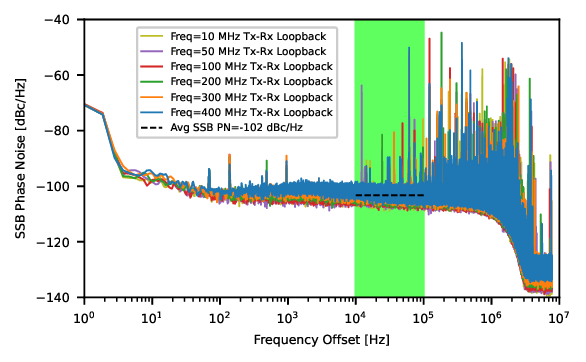}
\caption{SSB phase noise for selected tones after channelization. The shaded green region represents the possible region for Flux-Ramp Modulation (FRM) frequencies. The black dashed line represents the averaged noise level.}\label{phase_rx}
\end{figure}

On the receiver side, the frequency comb was amplified, down-converted to base-band and filtered by the RF-mixing board. Then, in the digital domain, the tones were channelized, down-converted, and filtered again. The IQ data streams of each channel were acquired, and the SSB phase noise density $\mathscr{L}(f)$ was calculated. Figure\,\ref{phase_rx} shows the phase noise of the same six reference tones measured in the previous step, but after being processed by the receiving chain.. A detailed analysis shows that the 1/f phase noise component was almost completely removed during down-conversion and sampling process, except at frequencies below 10\,Hz. This is consistent with the fact that all oscillators and clocks are locked to the same frequency reference maintaining strong coherence between transmitter and receiver for frequencies below 10\,kHz. Beyond our application, which is not sensitive to low-frequency noise, the noise below 10\,Hz it is being studied because of its importance for the readout of other types of detectors such as MKIDs\,\cite{Rantwijk}. In the case of frequencies above 1\,MHz the roll-off on the noise profile is due to the combination of the channelizer and a 1,6\,MHz Digital Down-Converter (DDC) low-pass filter while the phase noise plateau above 5\,MHz around -136\,dBc/Hz is consistent with the theoretical predictions for quantization noise. Within the band of interest, the noise profiles are dominated by the white-phase noise component and several spurious signals. The black dashed line in Figure\,\ref{phase_rx} corresponds to the average white phase noise of -102\,dBc/Hz and represents a degradation of less than 6\,dB with respect to the noise level present in the injected tones. The white phase noise component is a combination of the phase noise present in the injected tones and the thermal noise added by the RF-mixing board, while the spurious signals are mostly inter-modulation products produced by the non-linearities of the RF components in the receiver chain. Despite the intensity of the spurious signals, their impact can be reduced by narrowing the filtering stages and choosing the flux-ramp frequency carefully.

\section{Conclusion}\label{conc}

A novel SDR readout system has been developed and characterized, utilizing the third generation of RFSoC devices. A characterization methodology has been proposed that allows for evaluating the degradation in the readout performance under real operating conditions. The system was characterized within two scenarios: in transmission mode and in loopback mode. The readout noise was found to be lower than -100\,dBc/Hz for a multiplexing factor of about 1000 channels, being in good agreement with state-of-the-art readout systems. The equivalent noise represents an end-to-end degradation of less than 4\,dB with respect to the noise level imposed by the cryogenic amplifier. 
Finally, it was demonstrated that the RFSoC Gen3 devices are suitable enabling technologies for the next generation of readout systems for superconducting detectors. They will lower the costs, reduce the system complexity, and increase the system integration significantly.

\backmatter

%\bmhead{Supplementary information}

%If your article has accompanying supplementary file/s please state so here. 

%Authors reporting data from electrophoretic gels and blots should supply the full unprocessed scans for key as part of their Supplementary information. This may be requested by the editorial team/s if it is missing.

%Please refer to Journal-level guidance for any specific requirements.

\bmhead{Acknowledgments} 
Manuel García Redondo is supported by the Consejo Nacional de
Investigaciones Científicas y Técnicas (CONICET) as well as for the Helmholtz International Research School in Astroparticles and Enabling Technologies (HIRSAP).
Manuel García Redondo also acknowledges the support of the Karlsruhe School of Elementary and Astroparticle Physics: Science and Technology (KSETA).

%\section*{Declarations}

%Some journals require declarations to be submitted in a standardised format. Please check the Instructions for Authors of the journal to which you are submitting to see if you need to complete this section. If yes, your manuscript must contain the following sections under the heading `Declarations':

%\begin{itemize}
%\item Conflict of interest/Competing interests (check journal-specific guidelines for which heading to use)
%\item Ethics approval 
%\item Consent for publication
%\item Availability of data and materials
%\item Authors' contributions
%\end{itemize}

%\noindent
%If any of the sections are not relevant to your manuscript, please include the heading and write `Not applicable' for that section. 

\bibliography{sn-bibliography}% common bib file


\begin{thebibliography}{22}
\ifx \bisbn   \undefined \def \bisbn  #1{ISBN #1}\fi
\ifx \binits  \undefined \def \binits#1{#1}\fi
\ifx \bauthor  \undefined \def \bauthor#1{#1}\fi
\ifx \batitle  \undefined \def \batitle#1{#1}\fi
\ifx \bjtitle  \undefined \def \bjtitle#1{#1}\fi
\ifx \bvolume  \undefined \def \bvolume#1{\textbf{#1}}\fi
\ifx \byear  \undefined \def \byear#1{#1}\fi
\ifx \bissue  \undefined \def \bissue#1{#1}\fi
\ifx \bfpage  \undefined \def \bfpage#1{#1}\fi
\ifx \blpage  \undefined \def \blpage #1{#1}\fi
\ifx \burl  \undefined \def \burl#1{\textsf{#1}}\fi
\ifx \doiurl  \undefined \def \doiurl#1{\url{https://doi.org/#1}}\fi
\ifx \betal  \undefined \def \betal{\textit{et al.}}\fi
\ifx \binstitute  \undefined \def \binstitute#1{#1}\fi
\ifx \binstitutionaled  \undefined \def \binstitutionaled#1{#1}\fi
\ifx \bctitle  \undefined \def \bctitle#1{#1}\fi
\ifx \beditor  \undefined \def \beditor#1{#1}\fi
\ifx \bpublisher  \undefined \def \bpublisher#1{#1}\fi
\ifx \bbtitle  \undefined \def \bbtitle#1{#1}\fi
\ifx \bedition  \undefined \def \bedition#1{#1}\fi
\ifx \bseriesno  \undefined \def \bseriesno#1{#1}\fi
\ifx \blocation  \undefined \def \blocation#1{#1}\fi
\ifx \bsertitle  \undefined \def \bsertitle#1{#1}\fi
\ifx \bsnm \undefined \def \bsnm#1{#1}\fi
\ifx \bsuffix \undefined \def \bsuffix#1{#1}\fi
\ifx \bparticle \undefined \def \bparticle#1{#1}\fi
\ifx \barticle \undefined \def \barticle#1{#1}\fi
\bibcommenthead
\ifx \bconfdate \undefined \def \bconfdate #1{#1}\fi
\ifx \botherref \undefined \def \botherref #1{#1}\fi
\ifx \url \undefined \def \url#1{\textsf{#1}}\fi
\ifx \bchapter \undefined \def \bchapter#1{#1}\fi
\ifx \bbook \undefined \def \bbook#1{#1}\fi
\ifx \bcomment \undefined \def \bcomment#1{#1}\fi
\ifx \oauthor \undefined \def \oauthor#1{#1}\fi
\ifx \citeauthoryear \undefined \def \citeauthoryear#1{#1}\fi
\ifx \endbibitem  \undefined \def \endbibitem {}\fi
\ifx \bconflocation  \undefined \def \bconflocation#1{#1}\fi
\ifx \arxivurl  \undefined \def \arxivurl#1{\textsf{#1}}\fi
\csname PreBibitemsHook\endcsname

%%% 1
\bibitem[\protect\citeauthoryear{Abitbol et~al.}{2017}]{Abitbol2017}
\begin{botherref}
\oauthor{\bsnm{Abitbol}, \binits{M.H.}},
\oauthor{\bsnm{Ahmed}, \binits{Z.}},
\oauthor{\bsnm{Barron}, \binits{D.}},
\oauthor{\bsnm{Thakur}, \binits{R.B.}},
\oauthor{\bsnm{Bender}, \binits{A.N.}},
\oauthor{\bsnm{Benson}, \binits{B.A.}},
\oauthor{\bsnm{Bischoff}, \binits{C.A.}},
\oauthor{\bsnm{Bryan}, \binits{S.A.}},
\oauthor{\bsnm{Carlstrom}, \binits{J.E.}},
\oauthor{\bsnm{Chang}, \binits{C.L.}},
\oauthor{\bsnm{Chuss}, \binits{D.T.}},
\oauthor{\bsnm{Crowley}, \binits{K.T.}},
\oauthor{\bsnm{Cukierman}, \binits{A.}},
\oauthor{\bsnm{Haan}, \binits{T.}},
\oauthor{\bsnm{Dobbs}, \binits{M.}},
\oauthor{\bsnm{Essinger-Hileman}, \binits{T.}},
\oauthor{\bsnm{Filippini}, \binits{J.P.}},
\oauthor{\bsnm{Ganga}, \binits{K.}},
\oauthor{\bsnm{Gudmundsson}, \binits{J.E.}},
\oauthor{\bsnm{Halverson}, \binits{N.W.}},
\oauthor{\bsnm{Hanany}, \binits{S.}},
\oauthor{\bsnm{Henderson}, \binits{S.W.}},
\oauthor{\bsnm{Hill}, \binits{C.A.}},
\oauthor{\bsnm{Ho}, \binits{S.-P.P.}},
\oauthor{\bsnm{Hubmayr}, \binits{J.}},
\oauthor{\bsnm{Irwin}, \binits{K.}},
\oauthor{\bsnm{Jeong}, \binits{O.}},
\oauthor{\bsnm{Johnson}, \binits{B.R.}},
\oauthor{\bsnm{Kernasovskiy}, \binits{S.A.}},
\oauthor{\bsnm{Kovac}, \binits{J.M.}},
\oauthor{\bsnm{Kusaka}, \binits{A.}},
\oauthor{\bsnm{Lee}, \binits{A.T.}},
\oauthor{\bsnm{Maria}, \binits{S.}},
\oauthor{\bsnm{Mauskopf}, \binits{P.}},
\oauthor{\bsnm{McMahon}, \binits{J.J.}},
\oauthor{\bsnm{Moncelsi}, \binits{L.}},
\oauthor{\bsnm{Nadolski}, \binits{A.W.}},
\oauthor{\bsnm{Nagy}, \binits{J.M.}},
\oauthor{\bsnm{Niemack}, \binits{M.D.}},
\oauthor{\bsnm{O'Brient}, \binits{R.C.}},
\oauthor{\bsnm{Padin}, \binits{S.}},
\oauthor{\bsnm{Parshley}, \binits{S.C.}},
\oauthor{\bsnm{Pryke}, \binits{C.}},
\oauthor{\bsnm{Roe}, \binits{N.A.}},
\oauthor{\bsnm{Rostem}, \binits{K.}},
\oauthor{\bsnm{Ruhl}, \binits{J.}},
\oauthor{\bsnm{Simon}, \binits{S.M.}},
\oauthor{\bsnm{Staggs}, \binits{S.T.}},
\oauthor{\bsnm{Suzuki}, \binits{A.}},
\oauthor{\bsnm{Switzer}, \binits{E.R.}},
\oauthor{\bsnm{Tajima}, \binits{O.}},
\oauthor{\bsnm{Thompson}, \binits{K.L.}},
\oauthor{\bsnm{Timbie}, \binits{P.}},
\oauthor{\bsnm{Tucker}, \binits{G.S.}},
\oauthor{\bsnm{Vieira}, \binits{J.D.}},
\oauthor{\bsnm{Vieregg}, \binits{A.G.}},
\oauthor{\bsnm{Westbrook}, \binits{B.}},
\oauthor{\bsnm{Wollack}, \binits{E.J.}},
\oauthor{\bsnm{Yoon}, \binits{K.W.}},
\oauthor{\bsnm{Young}, \binits{K.S.}},
\oauthor{\bsnm{Young}, \binits{E.Y.}}:
{CMB-S4 Technology Book, First Edition}.
arXiv e-prints,
1706--02464
(2017)
\doiurl{10.48550/arXiv.1706.02464}
{\href{https://arxiv.org/abs/1706.02464}{{arXiv:1706.02464}}}
{[astro-ph.IM]}
\end{botherref}
\endbibitem

%%% 2
\bibitem[\protect\citeauthoryear{Hubmayr et~al.}{2018}]{Hubmayr_2018}
\begin{barticle}
\bauthor{\bsnm{Hubmayr}, \binits{J.}},
\bauthor{\bsnm{Austermann}, \binits{J.E.}},
\bauthor{\bsnm{Beall}, \binits{J.A.}},
\bauthor{\bsnm{Becker}, \binits{D.T.}},
\bauthor{\bsnm{Dober}, \binits{B.}},
\bauthor{\bsnm{Duff}, \binits{S.M.}},
\bauthor{\bsnm{Gao}, \binits{J.}},
\bauthor{\bsnm{Hilton}, \binits{G.C.}},
\bauthor{\bsnm{McKenney}, \binits{C.M.}},
\bauthor{\bsnm{Ullom}, \binits{J.N.}},
\bauthor{\bsnm{Lanen}, \binits{J.V.}},
\bauthor{\bsnm{Vissers}, \binits{M.R.}}:
\batitle{Low-temperature detectors for {CMB} imaging arrays}.
\bjtitle{Journal of Low Temperature Physics}
\bvolume{193}(\bissue{3-4}),
\bfpage{633}--\blpage{647}
(\byear{2018})
\doiurl{10.1007/s10909-018-2029-6}
\end{barticle}
\endbibitem

%%% 3
\bibitem[\protect\citeauthoryear{Geria et~al.}{2023}]{Geria2023}
\begin{barticle}
\bauthor{\bsnm{Geria}, \binits{J.M.}},
\bauthor{\bsnm{Hampel}, \binits{M.R.}},
\bauthor{\bsnm{Kempf}, \binits{S.}},
\bauthor{\bsnm{Bonaparte}, \binits{J.J.F.}},
\bauthor{\bsnm{Ferreyro}, \binits{L.P.}},
\bauthor{\bsnm{Redondo}, \binits{M.E.G.}},
\bauthor{\bsnm{Almela}, \binits{D.A.}},
\bauthor{\bsnm{Salum}, \binits{J.M.S.}},
\bauthor{\bsnm{Müller}, \binits{N.A.}},
\bauthor{\bsnm{Neira}, \binits{J.D.B.}},
\bauthor{\bsnm{Fuster}, \binits{A.E.}},
\bauthor{\bsnm{Platino}, \binits{M.}},
\bauthor{\bsnm{Etchegoyen}, \binits{A.}}:
\batitle{Suitability of magnetic microbolometers based on paramagnetic temperature sensors for cmb polarization measurements}.
\bjtitle{Journal of Astronomical Telescopes, Instruments, and Systems}
\bvolume{9}(\bissue{1}),
\bfpage{016002}
(\byear{2023})
\doiurl{10.1117/1.JATIS.9.1.016002} .
\bcomment{54.12.01; LK 01}
\end{barticle}
\endbibitem

%%% 4
\bibitem[\protect\citeauthoryear{Dober et~al.}{2021}]{Dober2020}
\begin{botherref}
\oauthor{\bsnm{Dober}, \binits{B.}},
\oauthor{\bsnm{Ahmed}, \binits{Z.}},
\oauthor{\bsnm{Arnold}, \binits{K.}},
\oauthor{\bsnm{Becker}, \binits{D.T.}},
\oauthor{\bsnm{Bennett}, \binits{D.A.}},
\oauthor{\bsnm{Connors}, \binits{J.A.}},
\oauthor{\bsnm{Cukierman}, \binits{A.}},
\oauthor{\bsnm{D'Ewart}, \binits{J.M.}},
\oauthor{\bsnm{Duff}, \binits{S.M.}},
\oauthor{\bsnm{Dusatko}, \binits{J.E.}},
\oauthor{\bsnm{Frisch}, \binits{J.C.}},
\oauthor{\bsnm{Gard}, \binits{J.D.}},
\oauthor{\bsnm{Henderson}, \binits{S.W.}},
\oauthor{\bsnm{Herbst}, \binits{R.}},
\oauthor{\bsnm{Hilton}, \binits{G.C.}},
\oauthor{\bsnm{Hubmayr}, \binits{J.}},
\oauthor{\bsnm{Li}, \binits{Y.}},
\oauthor{\bsnm{Mates}, \binits{J.A.B.}},
\oauthor{\bsnm{McCarrick}, \binits{H.}},
\oauthor{\bsnm{Reintsema}, \binits{C.D.}},
\oauthor{\bsnm{Silva-Feaver}, \binits{M.}},
\oauthor{\bsnm{Ruckman}, \binits{L.}},
\oauthor{\bsnm{Ullom}, \binits{J.N.}},
\oauthor{\bsnm{Vale}, \binits{L.R.}},
\oauthor{\bsnm{Winkle}, \binits{D.D.V.}},
\oauthor{\bsnm{Vasquez}, \binits{J.}},
\oauthor{\bsnm{Wang}, \binits{Y.}},
\oauthor{\bsnm{Young}, \binits{E.}},
\oauthor{\bsnm{Yu}, \binits{C.}},
\oauthor{\bsnm{Zheng}, \binits{K.}}:
A microwave squid multiplexer optimized for bolometric applications.
Applied Physics Letters
\textbf{118}
(2021)
\doiurl{10.1063/5.0033416}
\end{botherref}
\endbibitem

%%% 5
\bibitem[\protect\citeauthoryear{Rao et~al.}{2020}]{Rao_2020}
\begin{barticle}
\bauthor{\bsnm{Rao}, \binits{M.S.}},
\bauthor{\bsnm{Silva-Feaver}, \binits{M.}},
\bauthor{\bsnm{Ali}, \binits{A.}},
\bauthor{\bsnm{Arnold}, \binits{K.}},
\bauthor{\bsnm{Ashton}, \binits{P.}},
\bauthor{\bsnm{Dober}, \binits{B.J.}},
\bauthor{\bsnm{Duell}, \binits{C.J.}},
\bauthor{\bsnm{Duff}, \binits{S.M.}},
\bauthor{\bsnm{Galitzki}, \binits{N.}},
\bauthor{\bsnm{Healy}, \binits{E.}},
\bauthor{\bsnm{Henderson}, \binits{S.}},
\bauthor{\bsnm{Ho}, \binits{S.-P.P.}},
\bauthor{\bsnm{Hoh}, \binits{J.}},
\bauthor{\bsnm{Kofman}, \binits{A.M.}},
\bauthor{\bsnm{Kusaka}, \binits{A.}},
\bauthor{\bsnm{Lee}, \binits{A.T.}},
\bauthor{\bsnm{Mangu}, \binits{A.}},
\bauthor{\bsnm{Mathewson}, \binits{J.}},
\bauthor{\bsnm{Mauskopf}, \binits{P.}},
\bauthor{\bsnm{McCarrick}, \binits{H.}},
\bauthor{\bsnm{Moore}, \binits{J.}},
\bauthor{\bsnm{Niemack}, \binits{M.D.}},
\bauthor{\bsnm{Raum}, \binits{C.}},
\bauthor{\bsnm{Salatino}, \binits{M.}},
\bauthor{\bsnm{Sasse}, \binits{T.}},
\bauthor{\bsnm{Seibert}, \binits{J.}},
\bauthor{\bsnm{Simon}, \binits{S.M.}},
\bauthor{\bsnm{Staggs}, \binits{S.}},
\bauthor{\bsnm{Stevens}, \binits{J.R.}},
\bauthor{\bsnm{Teply}, \binits{G.}},
\bauthor{\bsnm{Thornton}, \binits{R.}},
\bauthor{\bsnm{Ullom}, \binits{J.}},
\bauthor{\bsnm{Vavagiakis}, \binits{E.M.}},
\bauthor{\bsnm{Westbrook}, \binits{B.}},
\bauthor{\bsnm{Xu}, \binits{Z.}},
\bauthor{\bsnm{Zhu}, \binits{N.}}:
\batitle{Simons observatory microwave squid multiplexing readout -- cryogenic rf amplifier and coaxial chain design}.
\bjtitle{Journal of Low Temperature Physics}
\bvolume{199},
\bfpage{807}--\blpage{816}
(\byear{2020})
\doiurl{10.1007/s10909-020-02429-y}
\end{barticle}
\endbibitem

%%% 6
\bibitem[\protect\citeauthoryear{Yu et~al.}{2022}]{Yu2022}
\begin{barticle}
\bauthor{\bsnm{Yu}, \binits{C.}},
\bauthor{\bsnm{Ahmed}, \binits{Z.}},
\bauthor{\bsnm{Frisch}, \binits{J.C.}},
\bauthor{\bsnm{Henderson}, \binits{S.W.}},
\bauthor{\bsnm{Silva-Feaver}, \binits{M.}},
\bauthor{\bsnm{Arnold}, \binits{K.}},
\bauthor{\bsnm{Brown}, \binits{D.}},
\bauthor{\bsnm{Connors}, \binits{J.}},
\bauthor{\bsnm{Cukierman}, \binits{A.J.}},
\bauthor{\bsnm{D'Ewart}, \binits{J.M.}},
\bauthor{\bsnm{Dober}, \binits{B.J.}},
\bauthor{\bsnm{Dusatko}, \binits{J.E.}},
\bauthor{\bsnm{Haller}, \binits{G.}},
\bauthor{\bsnm{Herbst}, \binits{R.}},
\bauthor{\bsnm{Hilton}, \binits{G.C.}},
\bauthor{\bsnm{Hubmayr}, \binits{J.}},
\bauthor{\bsnm{Irwin}, \binits{K.D.}},
\bauthor{\bsnm{Kuo}, \binits{C.-L.}},
\bauthor{\bsnm{Mates}, \binits{J.A.B.}},
\bauthor{\bsnm{Ruckman}, \binits{L.}},
\bauthor{\bsnm{Ullom}, \binits{J.}},
\bauthor{\bsnm{Vale}, \binits{L.}},
\bauthor{\bsnm{Winkle}, \binits{D.D.V.}},
\bauthor{\bsnm{Vasquez}, \binits{J.}},
\bauthor{\bsnm{Young}, \binits{E.}}:
\batitle{Slac microresonator rf (smurf) electronics: A tone-tracking readout system for superconducting microwave resonator arrays}.
\bjtitle{Review of Scientific Instruments}
\bvolume{94},
\bfpage{014712}
(\byear{2022})
\doiurl{10.1063/5.0125084}
\end{barticle}
\endbibitem

%%% 7
\bibitem[\protect\citeauthoryear{AMD}{2022}]{rfsoc_over}
\begin{botherref}
\oauthor{\bsnm{AMD}}:
Zynq UltraScale+ RFSoC Data Sheet: Overview.
(2022).
\url{https://docs.xilinx.com/v/u/en-US/ds889-zynq-usp-rfsoc-overview}
\end{botherref}
\endbibitem

%%% 8
\bibitem[\protect\citeauthoryear{AMD}{2023}]{rfsoc_ac}
\begin{botherref}
\oauthor{\bsnm{AMD}}:
Zynq UltraScale+ RFSoC Data Sheet: DC and AC Switching Characteristics (DS926).
(2023).
\url{https://docs.xilinx.com/r/en-US/ds926-zynq-ultrascale-plus-rfsoc}
\end{botherref}
\endbibitem

%%% 9
\bibitem[\protect\citeauthoryear{AMD}{2022}]{zcu216}
\begin{botherref}
\oauthor{\bsnm{AMD}}:
ZCU216 Evaluation Board User Guide (UG1390).
(2022).
\url{https://docs.xilinx.com/v/u/en-US/ug1390-zcu216-eval-bd}
\end{botherref}
\endbibitem

%%% 10
\bibitem[\protect\citeauthoryear{Sander et~al.}{2019}]{sander19}
\begin{barticle}
\bauthor{\bsnm{Sander}, \binits{O.}},
\bauthor{\bsnm{Karcher}, \binits{N.}},
\bauthor{\bsnm{Krömer}, \binits{O.}},
\bauthor{\bsnm{Kempf}, \binits{S.}},
\bauthor{\bsnm{Wegner}, \binits{M.}},
\bauthor{\bsnm{Enss}, \binits{C.}},
\bauthor{\bsnm{Weber}, \binits{M.}}:
\batitle{Software-defined radio readout system for the echo experiment}.
\bjtitle{IEEE Transactions on Nuclear Science}
\bvolume{66}(\bissue{7}),
\bfpage{1204}--\blpage{1209}
(\byear{2019})
\doiurl{10.1109/TNS.2019.2914665}
\end{barticle}
\endbibitem

%%% 11
\bibitem[\protect\citeauthoryear{Gartmann et~al.}{2022}]{Gartmann2022}
\begin{barticle}
\bauthor{\bsnm{Gartmann}, \binits{R.}},
\bauthor{\bsnm{Karcher}, \binits{N.}},
\bauthor{\bsnm{Gebauer}, \binits{R.}},
\bauthor{\bsnm{Krömer}, \binits{O.}},
\bauthor{\bsnm{Sander}, \binits{O.}}:
\batitle{Progress of the echo sdr readout hardware for multiplexed mmcs}.
\bjtitle{Journal of Low Temperature Physics}
\bvolume{209},
\bfpage{726}--\blpage{733}
(\byear{2022})
\doiurl{10.1007/s10909-022-02854-1}
\end{barticle}
\endbibitem

%%% 12
\bibitem[\protect\citeauthoryear{AMD}{2022}]{clk104}
\begin{botherref}
\oauthor{\bsnm{AMD}}:
CLK104 RF Clock Add-on Card User Guide (UG1437).
(2022).
\url{https://docs.xilinx.com/r/en-US/ug1437-clk104/Introduction}
\end{botherref}
\endbibitem

%%% 13
\bibitem[\protect\citeauthoryear{Karcher et~al.}{2020}]{Karcher2020}
\begin{barticle}
\bauthor{\bsnm{Karcher}, \binits{N.}},
\bauthor{\bsnm{Richter}, \binits{D.}},
\bauthor{\bsnm{Ahrens}, \binits{F.}},
\bauthor{\bsnm{Gartmann}, \binits{R.}},
\bauthor{\bsnm{Wegner}, \binits{M.}},
\bauthor{\bsnm{Krömer}, \binits{O.}},
\bauthor{\bsnm{Kempf}, \binits{S.}},
\bauthor{\bsnm{Enss}, \binits{C.}},
\bauthor{\bsnm{Weber}, \binits{M.}},
\bauthor{\bsnm{Sander}, \binits{O.}}:
\batitle{Sdr-based readout electronics for the echo experiment}.
\bjtitle{Journal of Low Temperature Physics}
\bvolume{200},
\bfpage{261}--\blpage{268}
(\byear{2020})
\doiurl{10.1007/s10909-020-02463-w}
\end{barticle}
\endbibitem

%%% 14
\bibitem[\protect\citeauthoryear{Ferreyro et~al.}{2023}]{Ferreyro2023}
\begin{barticle}
\bauthor{\bsnm{Ferreyro}, \binits{L.P.}},
\bauthor{\bsnm{Redondo}, \binits{M.G.}},
\bauthor{\bsnm{Hampel}, \binits{M.R.}},
\bauthor{\bsnm{Almela}, \binits{A.}},
\bauthor{\bsnm{Fuster}, \binits{A.}},
\bauthor{\bsnm{Salum}, \binits{J.}},
\bauthor{\bsnm{Geria}, \binits{J.M.}},
\bauthor{\bsnm{Bonaparte}, \binits{J.}},
\bauthor{\bsnm{Bonilla-Neira}, \binits{J.}},
\bauthor{\bsnm{Müller}, \binits{N.}},
\bauthor{\bsnm{Karcher}, \binits{N.}},
\bauthor{\bsnm{Sander}, \binits{O.}},
\bauthor{\bsnm{Platino}, \binits{M.}},
\bauthor{\bsnm{Weber}, \binits{M.}},
\bauthor{\bsnm{Etchegoyen}, \binits{A.}}:
\batitle{An implementation of a channelizer based on a goertzel filter bank for the read-out of cryogenic sensors}.
\bjtitle{Journal of Instrumentation}
\bvolume{18}(\bissue{06}),
\bfpage{06009}
(\byear{2023})
\doiurl{10.1088/1748-0221/18/06/P06009}
\end{barticle}
\endbibitem

%%% 15
\bibitem[\protect\citeauthoryear{AMD}{2023}]{rfdc}
\begin{botherref}
\oauthor{\bsnm{AMD}}:
Zynq UltraScale+ RFSoC RF Data Converter V2.6 Gen 1/2/3/DFE LogiCORE IP Product Guide (PG269).
(2023).
\url{https://docs.xilinx.com/r/en-US/pg269-rf-data-converter}
\end{botherref}
\endbibitem

%%% 16
\bibitem[\protect\citeauthoryear{Ahrens}{2022}]{ahrensthesis}
\begin{botherref}
\oauthor{\bsnm{Ahrens}, \binits{F.K.}}:
Cryogenic read-out system and resonator optimisation for the microwave squid multiplexer within the echo experiment.
PhD thesis,
Heidelberg University
(2022).
\doiurl{10.11588/heidok.00032038}
\end{botherref}
\endbibitem

%%% 17
\bibitem[\protect\citeauthoryear{Mates et~al.}{2012}]{Mates2012}
\begin{barticle}
\bauthor{\bsnm{Mates}, \binits{J.A.B.}},
\bauthor{\bsnm{Irwin}, \binits{K.D.}},
\bauthor{\bsnm{Vale}, \binits{L.R.}},
\bauthor{\bsnm{Hilton}, \binits{G.C.}},
\bauthor{\bsnm{Gao}, \binits{J.}},
\bauthor{\bsnm{Lehnert}, \binits{K.W.}}:
\batitle{Flux-ramp modulation for squid multiplexing}.
\bjtitle{Journal of Low Temperature Physics}
\bvolume{167},
\bfpage{707}--\blpage{712}
(\byear{2012})
\doiurl{10.1007/s10909-012-0518-6}
\end{barticle}
\endbibitem

%%% 18
\bibitem[\protect\citeauthoryear{Salum et~al.}{2023}]{Salum2023}
\begin{barticle}
\bauthor{\bsnm{Salum}, \binits{J.M.}},
\bauthor{\bsnm{Muscheid}, \binits{T.}},
\bauthor{\bsnm{Fuster}, \binits{A.}},
\bauthor{\bsnm{{Garcia Redondo}}, \binits{M.E.}},
\bauthor{\bsnm{Hampel}, \binits{M.R.}},
\bauthor{\bsnm{Ferreyro}, \binits{L.P.}},
\bauthor{\bsnm{Geria}, \binits{J.M.}},
\bauthor{\bsnm{Bonilla-Neira}, \binits{J.}},
\bauthor{\bsnm{Müller}, \binits{N.}},
\bauthor{\bsnm{Bonaparte}, \binits{J.}},
\bauthor{\bsnm{Almela}, \binits{A.}},
\bauthor{\bsnm{Ardila-Perez}, \binits{L.E.}},
\bauthor{\bsnm{Platino}, \binits{M.}},
\bauthor{\bsnm{Sander}, \binits{O.}},
\bauthor{\bsnm{Weber}, \binits{M.}}:
\batitle{Aliasing effect on flux ramp demodulation: Nonlinearity in the microwave squid multiplexer}.
\bjtitle{Journal of Low Temperature Physics}
\bvolume{213},
\bfpage{223}--\blpage{236}
(\byear{2023})
\doiurl{10.1007/s10909-023-02993-z}
\end{barticle}
\endbibitem

%%% 19
\bibitem[\protect\citeauthoryear{Schuster et~al.}{2023}]{Schuster_2023}
\begin{barticle}
\bauthor{\bsnm{Schuster}, \binits{C.}},
\bauthor{\bsnm{Wegner}, \binits{M.}},
\bauthor{\bsnm{Kempf}, \binits{S.}}:
\batitle{Simulation framework for microwave {SQUID} multiplexer optimization}.
\bjtitle{Journal of Applied Physics}
\bvolume{133}(\bissue{4}),
\bfpage{044503}
(\byear{2023})
\doiurl{10.1063/5.0135124}
\end{barticle}
\endbibitem

%%% 20
\bibitem[\protect\citeauthoryear{Shibasaki et~al.}{2020}]{Shibasaki}
\begin{bchapter}
\bauthor{\bsnm{Shibasaki}, \binits{Y.}},
\bauthor{\bsnm{Asami}, \binits{K.}},
\bauthor{\bsnm{Aoki}, \binits{R.}},
\bauthor{\bsnm{Hatta}, \binits{A.}},
\bauthor{\bsnm{Kuwana}, \binits{A.}},
\bauthor{\bsnm{Kobayashi}, \binits{H.}}:
\bctitle{Analysis and design of multi-tone signal generation algorithms for reducing crest factor}.
In: \bbtitle{2020 IEEE 29th Asian Test Symposium (ATS)},
pp. \bfpage{1}--\blpage{6}
(\byear{2020}).
\doiurl{10.1109/ATS49688.2020.9301549}
\end{bchapter}
\endbibitem

%%% 21
\bibitem[\protect\citeauthoryear{Rubiola}{2010}]{rubiola2010phase}
\begin{bbook}
\bauthor{\bsnm{Rubiola}, \binits{E.}}:
\bbtitle{Phase Noise and Frequency Stability in Oscillators}.
\bsertitle{The Cambridge RF and Microwave Engineering Series}.
\bpublisher{Cambridge University Press},
\blocation{Cambridge}
(\byear{2010})
\end{bbook}
\endbibitem

%%% 22
\bibitem[\protect\citeauthoryear{van Rantwijk et~al.}{2016}]{Rantwijk}
\begin{barticle}
\bauthor{\bsnm{Rantwijk}, \binits{J.}},
\bauthor{\bsnm{Grim}, \binits{M.}},
\bauthor{\bsnm{Loon}, \binits{D.}},
\bauthor{\bsnm{Yates}, \binits{S.}},
\bauthor{\bsnm{Baryshev}, \binits{A.}},
\bauthor{\bsnm{Baselmans}, \binits{J.}}:
\batitle{Multiplexed readout for 1000-pixel arrays of microwave kinetic inductance detectors}.
\bjtitle{IEEE Transactions on Microwave Theory and Techniques}
\bvolume{64}(\bissue{6}),
\bfpage{1876}--\blpage{1883}
(\byear{2016})
\doiurl{10.1109/TMTT.2016.2544303}
\end{barticle}
\endbibitem

\end{thebibliography}
%\bibliographystyle{abbrvnat}
%\input sn-article.bbl

\end{document}